# New Parameter in Polymer - Assisted Sol - Gel Deposition

Pylnev M.A.

In this work a new parameter for the Polymer Assisted Sol- Gel Deposition method is proposed. It is shaking of the sol stabilized by the polymer prior to the deposition. It was found in attempt to decrease the amount of [111] - oriented $CeO_2$ grains on a monocrystalline substrate $SrTiO_3$ (001). It was found that the shaking decreases this amount significantly.

**Introduction**

The critical current of superconducting tape based on the Ni – $YBa_2Cu_3O_7$ ensemble, is in direct proportion to the amount of [001] - oriented $YBa_2Cu_3O_7$ (YBCO) grains on Ni surface. An increase of this amount is achieved by using buffer layers between the Ni and YBCO layers. In this paper, a new regularity of the growth of the buffer layer $CeO_2$ in the Polymer Assisted Sol- Gel Deposition method is proposed. This buffer has a very suitable lattice parameter and therefore it is used as a cap layer in the buffer architecture. The main problem of using cerium dioxide as a buffer is the growth of non- epitaxial to the substrate $CeO_2(111)$ grains [1,2] that do not encourage the growth of YBCO(001) grains.

The following growth laws of non- epitaxial to the substrate $CeO_2(111)$ grains were found in literature. In [3], $CeO_2$ films are obtained by the Metalorganic Deposition (MOD) method and the reason of formation of $CeO_2(111)$ grains is considered that they nucleate far away from the substrate where there is not a motive to orient in the [001] direction. In contrast, in [4], the Pulsed Laser Deposition method was used for deposition of $CeO_2$ films. In the paper, it was proposed that the reason of formation of $CeO_2(111)$ grains could be $Ce_2O_3$ phase that is obtained in an oxygen- less environment: "the minimal mismatch between the lattice constants of $CeO_2$ and $Ce_2O_3$ is occurred when $CeO_2$ (111) is parallel to $Ce_2O_3(0001)$". However, the authors of [5], who prepared $CeO_2$ films by sol- gel method, suggested that the reason of formation of $CeO_2(111)$ grains is "local element disorders or lattice deformations suppressing epitaxial growth (with orientation [001])) with development of the more stable [111] orientation as a result".

In the present work a new parameter of the Polymer Assisted Sol- Gel Deposition method is introduced. As it will be shown later in this work, shaking the sol stabilized by the polymer prior to the deposition influences the growth of the films. So, it may provide deeper understanding of mechanisms of films growth in this deposition process.

**Experiment**

In the present work $CeO_2$ films with thickness of approximately 30 nm were deposited by the Polymer Assisted Sol- Gel Deposition method on monocrystalline $SrTiO_3$(001) substrates with a roughness of 1 - 2 nm.

The preparation method of the sol can be found in [6]. A polymer polyvinylcaprolactam (PVCL) is added into the sol at the stage of its production for: providing uniform film formation by adhesive and amphiphilic properties of the polymer [7 - 9], inhibiting agglomeration of the particles in the solution [7] and preventing formation of droplets on the substrate [8]. The properties of the films were different if the sols stabilized by the polymers had been shaken. Now the process of the film deposition will be described. The sol stabilized by the polymer is deposited on a substrate with a defined amount by pouring it into a flat cuvette with the substrate using a pipette (Labnet). Then this system is dried in a moisture analyzer (AND MF-50) under 50°C until 100% of humidity has been lost. After this stage, the sol stabilized by the polymer is transformed to a cotton- like polymer system, which pores contain $CeO_2$ nanoparticles [6]. Then, the ensemble of substrate and the described system (hereinafter the specimen) is placed into a quartz reactor. Then the reactor is placed to a dental oven (MIMP- V) where it is annealed for a defined period of time, in defined temperature regime and gas flow through the reactor. After this, the compartment of the oven is put down that provides quenching of specimen to room temperature.

In this study, three different specimens were prepared as follows:
1) Specimen 1 was obtained without using the shaking. The sol stabilized by the polymer was not shaken. Before the deposition, the sol stabilized by the polymer was stored in air for approximately 3 hours until $Ce(OH)_3$ completely transformed to $CeO_2$; the amount of $Ce^{3+}$ ions in the sol stabilized by the polymer was controlled by the Perkin Elmer Lambda 9 UV/VIS/NIR spectrophotometer. Specimen was annealed for an hour under 1000ºC in a gas mixture of 95% Ar + 5% $H_2$ with a flow rate of 10 l/h. Quenching of the specimen was performed from the temperature of ~ 900 °C.
2) For specimen 2, the sol stabilized by the polymer, after being stored in air, was shaken for ca. 2 - 3 minutes. The parameters of annealing were the same as for the specimen 1.
3) Specimen 3 was annealed in static air for 60 minutes at 1000ºC. The rest parameters were equal to the parameters of the specimen 1.

The crystal structures of samples were characterized by powder X-ray Diffraction (XRD) (Bourevestnik Dron – 3) using Cu Kα radiation. The surface morphologies of the specimens were investigated using Atomic Force Microscopy (AFM)(Femtoscan, Advanced Technologies Center). The measurements were performed in contact mode.

**Results**

X- ray diffraction patterns (XRD patterns) of the specimens 1, 2 and 3 are shown in Fig. 1. It should be noted that the ratio of the square of the structure amplitude for the reflection from the $CeO_2(111)$ planes to that for the second order reflection from the $CeO_2(100)$ planes is $F_{111}^2/F_{200}^2 \sim 2$. Hence, the ratio of [111] - oriented $CeO_2$ grains is less than it follows from the XRD patterns. It can be seen from the XRD patterns that the ratio of the intensity of the (200) peak to the intensity of the (111) peak $R = I_{200}/I_{111}$ increases if the sol stabilized by the polymer has been shaken (from R = 1.7 for the specimen 1 to R = 5.76 for specimen 2). AFM images of surface of the specimens are shown in Fig. 2. It can be seen that the surface of the specimen 2 (Fig. 2, 2) more blurred than the surface of the specimen 1 and the grains of the specimen 2 are much less delineated (Fig. 2, 1).

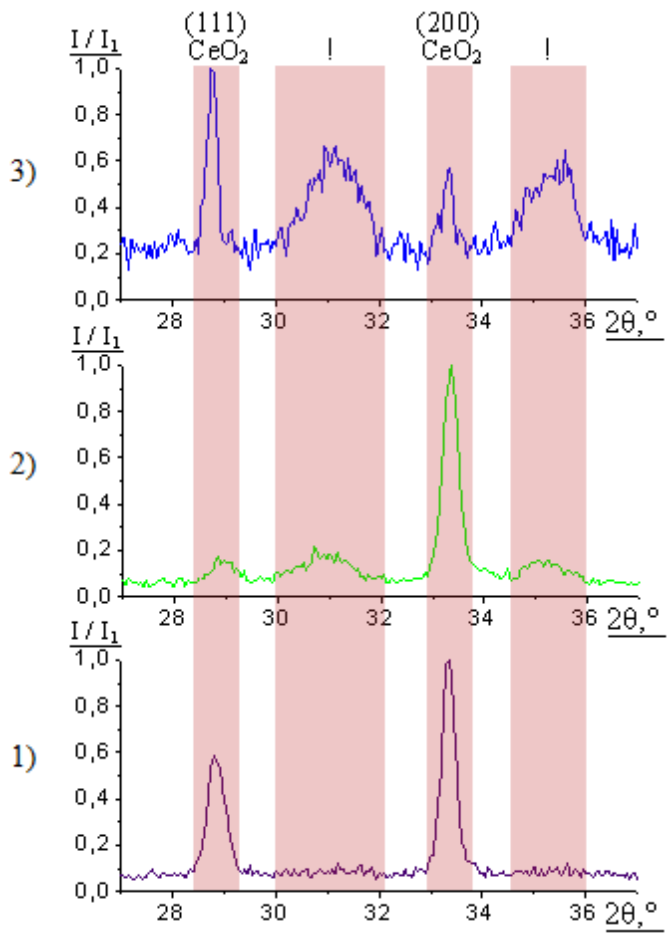

Fig.1.  XRD patterns of the CeO$_2$ films on SrTiO$_3$:
1) Specimen 1 that was obtained without using the shaking;
2) Specimen 2 that was obtained with using the shaking;
3) Specimen 3 that was annealed in static air.
The patterns have been normalized for clarity.

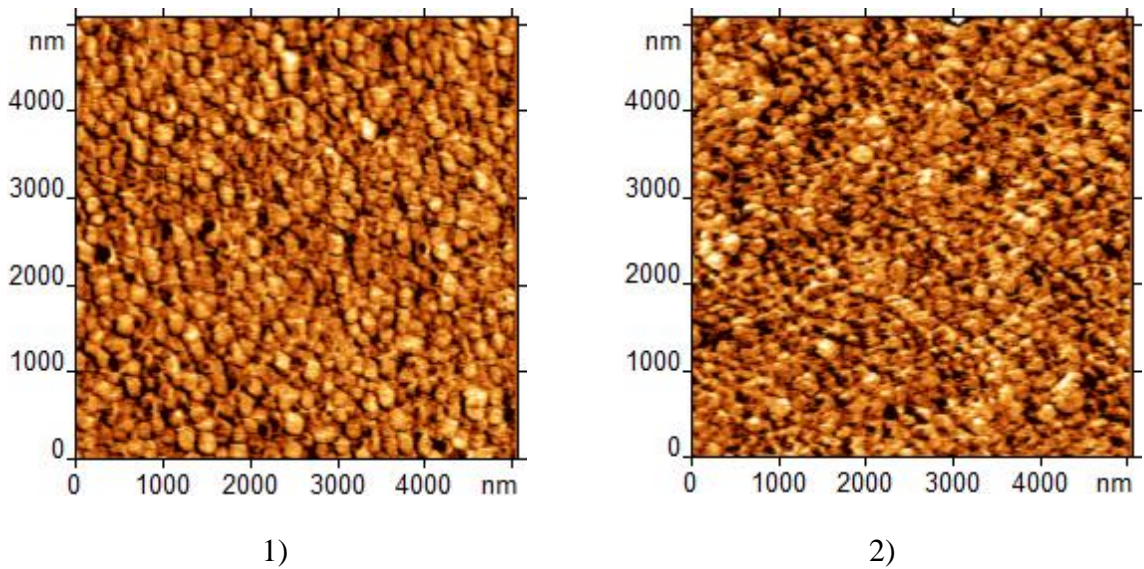

Fig. 2. AFM images of the surfaces of the specimen 1 obtained without using the shaking (1) and the specimen 2 (2)

It must be noticed that there are unexpected peaks at $2\theta = 31°$ and $2\theta = 35°$ on the XRD patterns corresponding to the specimen 2 (Fig. 1, 2). Presumably, these peaks belong to adducts of the reaction between $CeO_2$ and the products of the polymer decomposition. Moreover, the peaks are weaker but broader than the peaks of $CeO_2(100)$ and $CeO_2(111)$ grains. That means that these peaks correspond to smaller amounts of poor crystalline phases. Furthermore, the peaks with the same positions are seen on the XRD pattern of the specimen 3 that was annealed in static air (Fig. 1, 3).

It may need to be noted that the presented specimen 2 is an average among many. The shaking parameter is still not clearly understood however there were specimens obtained with using the shaking showing no presence of $CeO_2(111)$ grains at all.

**Conclusion**

In this work a new parameter for the Polymer Assisted Sol- Gel Deposition method is proposed. It is shaking of the sol stabilized by the polymer prior to the deposition. It was shown that if the stabilized by polymer sol was shaken, the ratio of $CeO_2(111)$ grains to $CeO_2(001)$ grains obtained on $SrTiO_3(001)$ substrate decreases significantly. In addition, it changes the surface of the $CeO_2$ films making grains much less delineated.